\documentclass[twocolumn,preprintnumbers,amsmath,amssymb,superscriptaddress]{revtex4}

\usepackage[dvips]{epsfig}

\usepackage{float}

\begin{document}

\title{Comment on ``Time reversal polarization and a $Z_2$ adiabatic spin pump''}
\author{Yi-Dong Wu}
\affiliation{Department of Applied Physics, School of Science, Yanshan University, Qinhuangdao, Hebei, 066004, China}
\email{wuyidong@ysu.edu.cn}

\begin{abstract}
In Ref \onlinecite{Fu2006}[Phy. Rev. B 74, 195312(2006)] Fu and Kane propose a spin pump for one-dimensional(1D) insulating Hamiltonians. They claim that this spin pump is a ``$Z_2$ pump'' because ``For an isolated system, a single closed cycle of the pump changes the expectation value of the spin at each end even when spin-orbit interactions violate the conservation of spin. A second cycle, however, returns the system to its original state''. A $Z_2$ topological invariant is proposed to characterize the ``$Z_2$ pump''. In this comment we show their discussion on the spin pump is inaccurate. Their reason why the isolated system return to its original state after second cycle is unjustified and several claims contradict to this return of the system are made in Ref \onlinecite{Fu2006}. Detailed calculations and concrete examples show the degeneracy of the first excited state at $t=0,T,...$ is not split by the electron-electron interaction in the way described in Ref \onlinecite{Fu2006} and there is level crossing at $t=T$. In fact, despite of a detailed search, not a single system behave as described in Fig. 1(d) in Ref \onlinecite{Fu2006} has been found. Thus we conclude the isolated system won't return to its original state after two cycles and the spin pump is not a ``$Z_2$ pump'' in general.
\end{abstract}

\pacs{73.43.-f, 72.25.Hg, 75.10.Pq, 85.75.-d}
\maketitle
In Ref \onlinecite{Fu2006} a spin pump for one-dimensional(1D) insulating Hamiltonians is proposed. By studying the evolution of the end state of a tight-binding model they claim the spin pump is a ``$Z_2$ pump'' because the system will return to its original state after two cycles. The subsequent discussions are all based on this conclusion. $Z_2$ topological invariant is proposed to characterize the ``$Z_2$ pump''. Because the $Z_2$ pump can not pump spin in isolated system, they discuss the possibility that the spin pump may pump spin when connected to the reservoirs. If the spin pump fail to be a ``$Z_2$ pump'', all those discussions  will be groundless. For example there is no reason to use a $Z_2$ invariant to characterize a ``non-$Z_2$ pump''. Another topological invariant must be introduced to characterize the spin pump. In this comment we show the spin pump is not a ``$Z_2$ pump'' in general.\\
We first point out a serious technical flaw in the discussion of the evolution of the end states in Ref \onlinecite{Fu2006}. In Fig. 1(d) in Ref \onlinecite{Fu2006}, they plot for $0<t<2T$ the energies of the lowest few many-body eigenstates associated with a single end, obtained by considering particle-hole excitations built from the single-particle states localized at that end. The energies of the particle-hole excitations of end states in the figure are continuous functions of $t$. However, a simple analysis shows it is impossible. As shown in the figure, the ground state of the end at $t=0$ is nondegenerate and the ground state at $t=T/2$ has two-fold Kramers degeneracy, which means there are even number of electrons at $t=0$ and odd at $t=T/2$ at the end. So there must be a discontinuous change of the electron number between $t=0$ and $t=T/2$. The energy of the end will also experience a discontinuous change in this interval. The mistake in the Fig. 1(d) in Ref \onlinecite{Fu2006} is that the authors fail to take into account the exchange of the electrons between the end and the bulk insulator.\\

\begin{figure}
  \includegraphics[width=3.5in,clip=true]{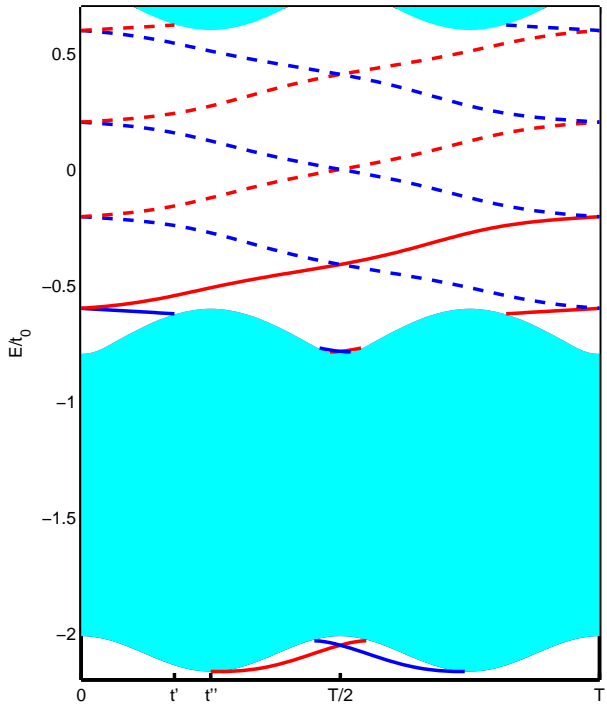}\\
  \caption{The energy spectrum of the tight-binding model in Ref \onlinecite{Fu2006}. The two branches of time-reversal related end states are denoted by different colors. The solid lines and doted lines represent occupied and unoccupied end states respectively. }\label{p1}
\end{figure}
To make our point we study a cycle of the spin pump in detail. We consider a 1000-sites chain instead of a 24-site chain with 12 extra sites added at each end because if we use a 24-site chain sometimes it's impossible to differentiate the end states and the bulk states. Fig. 1 shows the energy spectrum of the chain as functions of $t$. By calculating the average positions and the mean square deviations from the average positions we establish that only the cross lines in Fig .1 represent localized states at the end, other states are extended and can be considered as belong to the bulk bands. Now we study the adiabatic evolution of one end. We consider the case that two electrons occupy the lowest energy level at the end at $t=0$. Fig. 1 shows that the number of the electrons at the end $n$ is constantly changing in this adiabatic process, e.g. $n=2$ for $0<t<t'$, $n=1$ for $t'<t<t''$ and $n=5$ at $t=T/2$. Obviously, the variation of the number of the electrons at the end is due to the fact that the electrons merge into the bulk or electrons come out of the bulk and become localized at the end. So the end of the chain is an open system, it can exchange electrons and energy with the bulk even when the system is not connected with reservoirs. The energy of the particle-hole excitations built from the single-particle states localized at that end is the sum of the energies of the occupied electrons at the end.  It's clearly a discontinuous function of $t$. Though this mistake do not affect whether the spin pump is $Z_2$ or not, it does reflect that the discussions in Ref \onlinecite{Fu2006} lack a true understanding of how the spin pump works.\\

Now we focus on whether the spin pump is a ``$Z_2$ pump'' or not. First we assume the single-electron picture is valid. In the single-electron picture the topology of the electron bands ensure that there are two branches of time-reversal related gapless end states as illustrated Fig.1. Though the exchange of electrons between end and the bulk is complex, the net result of one cycle is simple, one electron at the end merges into the bulk and another electron comes out of the bulk and becomes localized at the end as illustrated in Fig 1. Those two electrons are in time-reversal related states and belong to different branches of the end states. Then, after a cycle one electron is added to one of the two branches end states and a hole is added to another. As illustrated in Fig.1 in the process of adiabatic evolution two branches of time reversal related states don't mix, so the added electron don't fill the hole. Clearly after $N$ cycles there will be $N$ electrons added to one branch of end states and $N$ holes to another provided there are enough end states to put them and the system won't return to its original state after two cycles. So if the single-electron picture is valid the spin pump is definitely not a ``$Z_2$ pump''.\\

In Ref \onlinecite{Fu2006} the authors claim that electron-electron interaction will destroy the single-electron picture completely and the system will return to the original state after two cycles.  In their argument the return of the system to the origin state after the second cycle relies on the fact that the four-fold degeneracy of the first excited state at the end when $t=T$ is split by the electron-electron interaction and there is no level crossing at $t=T$  as shown in FIG. 1.(d) in Ref \onlinecite{Fu2006}. The whole ``$Z_2$ pump'' theory depends on this degeneracy splitting. However, the authors give no justification besides a statement that ``Electron-electron interactions, however, will in general split this degeneracy, as shown in the inset(FIG. 1.(d))''. No argument, example or reference are provided to support the this conclusion. If this kind of degeneracy splitting do not happen the $Z_2$ pump theory will certainly fail. So the degeneracy splitting is vital to the success of the ``$Z_2$ pump'' theory.\\

Before discussing whether the electron-electron interaction can split the degeneracy or not we must point out even the authors of Ref \onlinecite{Fu2006} do not realize the consequence of this degeneracy splitting, because they make several claims that contradict to this conclusion e.g. ``For an isolated system, a single closed cycle of the pump changes the expectation value of the spin at each end even when spin-orbit interactions violate the conservation of spin.'',``In one cycle [...] (though the expectation value of the spin at the end changes by a nonquantized amount)'' and ``the eigenstates before and after adiabatic flux insertion [...] cannot be connected by any local time reversal invariant operator.''.\\
 If the four-fold degeneracy is split as in FIG. 1.(d) in Ref \onlinecite{Fu2006} none of those claims will be true. The proof of our conclusion is very simple. At $t=T$ Hamiltonian of the end is time-reversal invariant. If one state is an eigenstate of the Hamiltonian, the time-reversed state will be an eigenstate with same energy. If the first exited state of the end $|u_1\rangle$ is nondegenerate as shown in FIG. 1.(d) in Ref \onlinecite{Fu2006}, it must be time-reversal symmetric, that is $\hat{T}|u_1\rangle=c|u_1\rangle$, $\hat{T}$ is time-reversal operator and $c$ is a phase factor. We can choose $c=1$ by multiplying a constant to $|u_1\rangle$. We stick to this choice in following discussions. The same result applies to the nondegenerate ground state $|u_0\rangle$ at $t=0,T...$. Then we have $\hat{T}|u_1\rangle=|u_1\rangle$ and $\hat{T}|u_0\rangle=|u_0\rangle$. However, spin operator is odd under time-reversal transformation. Thus the average spin at the end before and after one close cycle are both zeros. So the expectation value of the spin at the end won't change if the four-fold degeneracy is split as in FIG. 1.(d) in Ref \onlinecite{Fu2006}.\\
 Because $|u_0\rangle$ and $|u_1\rangle$ describe the states of the end, they are both localized. Clearly the operator $|u_1\rangle\langle u_0|+|u_1\rangle\langle u_0|$ is a local time-reversal invariant Hermite operator that connect the eigenstates before and after the cycle.\\
 On the other hand, if the expectation value of spin does change and becomes nonzero after one cycle, the eigenstate after one cycle $|u_1\rangle$ must has at least two-fold degeneracy , and there will be level crossing at $t=T$. This is easy to understand. $|u_1\rangle$ and $\hat{T}|u_1\rangle$ must have opposite nonzero average spins, so they are linearly independent. Because the Hamiltonian is time-reversal invariant, $|u_1\rangle$ and $\hat{T}|u_1\rangle$ are eigenstates with same energy. Then the energy is at least two-fold degenerate. Because of this level crossing the end won't return to it's origin state after two cycles.\\

 Those self-contradictory results cast doubt on degeneracy splitting of the first excited state at $t=0,T$, which, they claim, is the general situation in Ref \onlinecite{Fu2006}. In fact, if there is no spin-obit coupling and the spin is conserved at the end there will always be a three-fold degeneracy due to the spin triplet. Thus if spin-obit coupling of the added atoms is weak enough to be neglected, the spin pump won't be a $Z_2$ pump. So even if the degeneracy is split as described in Ref \onlinecite{Fu2006}, it is by the spin-obit coupling at the end, not by the electron-electron interaction as claimed in Ref \onlinecite{Fu2006}.\\
 The problem is whether the spin-obit coupling can split the degeneracy or not. After a comprehensive search, we can not find a single system behaves the way as described in Ref \onlinecite{Fu2006}. On the contrary, we find plenty of examples where the degeneracy due to four ways of making particle-hole excitations with two pairs of Kramers degenerate states is not all split by the electron-electron interaction\cite{Nowak,Golovach,Fasth,Climente,Chakraborty,Pietil}.\\
  In Fig. 1 there are two electrons localized at the end 1D chain at $t=T$. The two-electron quantum dots in zero magnetic field are good analogy of the end of the chain. Without the electron-electron interaction the first excited stated in a quantum dot will be four-fold degenerate due to four ways of making particle-hole excitations with two pairs of Kramers degenerate states\cite{Nowak}. Though the singlet-triplet are split by the electron-electron interaction, the degeneracy within the triplet(first excited state) remains even with spin-obit interaction in various types of two-electron quantum dots\cite{Nowak,Golovach,Fasth,Climente,Chakraborty,Pietil}. It's highly possible that the first excited state of the end has the same degeneracy as the first exited states of the quantum dots.\\
  To further confirm our conclusion we consider electron-electron interaction at the end the chain with the tight-binding model in Ref \onlinecite{Fu2006}. The single-electron Hamiltonian of the end can be written as
\begin{equation}
\hat{H}_0=\sum_n\varepsilon_n \hat{\psi}_n^\dag\hat{\psi}_n
\end{equation}
where $\hat{\psi}_n^\dag$ and $\hat{\psi}_n$ are the the creation and annihilation operators of the electrons at the single-electron end states $|\psi_n\rangle=\hat{\psi}^\dag_n|0\rangle$. $\varepsilon_n$ is the eigenenergy of the single-electron end states. $|\psi_n\rangle$ and $\varepsilon_n$ are obtained by diagonalization the non-interacting Hamiltonian. In this tight-binding model only one space freedom is considered at each site, the interaction between electrons can be expressed as
\begin{equation}
\hat{H}_1=\frac{1}{2}\sum_{iji'j'}\sum_{\alpha\beta}U_{iji'j'}c^\dag_{i\alpha}c^\dag_{j\beta}c_{j'\beta}c_{i'\alpha}
\end{equation}
where $U_{iji'j'}$ is the matrix element of the Coulomb interaction in the tight-binding representation
\begin{equation}
\begin{split}
  U_{iji'j'}=&\int\varphi^*(x_1-X_i)\varphi^*(x_2-X_j)\\
(e^2/|x_1-x_2|)&\varphi(x_1-X_i')\varphi(x_2-X_{j'})dx_1dx_2
  \end{split}
\end{equation}

$\varphi(x_1-X_{i})$ is the space Wannier function and $X_{i}$ is the position of the site(in unit of lattice constant). The two freedoms(with different spin) of one site are related by the time-reversal operator, the space Wannier function can be chosen to be real. With this choice all the matrix elements are real. Since we don't have much information about the Wannier function, we must make some approximations before further calculation. The most commonly used approximation is the Hubbard model, in which only those matrix elements involving the same site ($i=i'=j=j'$) are retained. For our purpose we also consider the interactions between electrons at different site ($i=i',j=j'\neq i$). The matrix elements of these terms must be approximated. We approximate the interactions between electrons at different sites by both the long range type
\begin{eqnarray}
\begin{split}
 U_{iiii}=&U \\
U_{ijij}=&U_1/|X_i-X_j|(i\neq j) \\
\end{split}
\end{eqnarray}
and the screened type

\begin{equation}
  U_{ijij}=U'_1\exp(-\alpha|X_i-X_j|)/|X_i-X_j| (i\neq j)\\
\end{equation}
Other terms are neglected because they will scatter the electrons in or out of the end, which makes the number of electrons occupying the end indefinite(also because they are relatively small). We assume the occupation number of the end states is constant, otherwise  the end can not be considered as a closed system even for a given $t$, that is the end cannot be considered separately. And if the electrons are constantly scattered in and out of the end, the energy of particle-hole excitations at the end will become meaningless.\\
 The calculation method is similar to that is used in Ref \onlinecite{Nowak}. The Hamiltonian is is diagonalized in a basis constructed of single-electron states. We calculate the eigenenergies for $T-1<t<T+1$ where there are eight end states for a given end. Three cases with two, four and six occupied electrons at the end are considered. The bases are $\hat{\psi}^\dag_m\hat{\psi}^\dag_n|0\rangle(m<n)$, $\hat{\psi}^\dag_m\hat{\psi}^\dag_n\hat{\psi}^\dag_o\hat{\psi}^\dag_p|0\rangle(m<n<o<p)$ and $\hat{\psi}^\dag_m\hat{\psi}^\dag_n\hat{\psi}^\dag_o\hat{\psi}^\dag_p\hat{\psi}^\dag_q\hat{\psi}^\dag_r|0\rangle(m<n<o<p<q<r)$ respectively.\\
\begin{figure}
  \includegraphics[width=3.5in,clip=true]{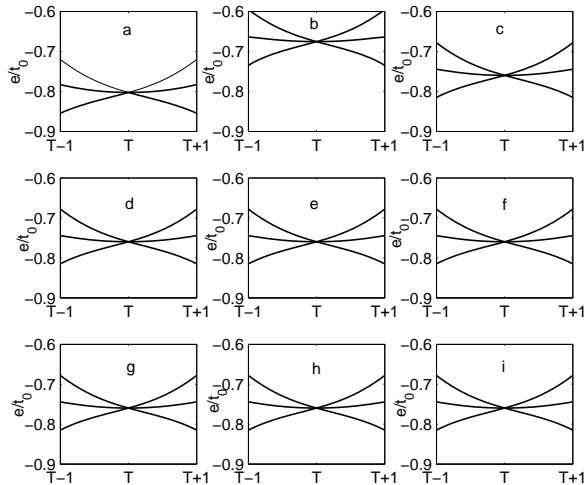}\\
  \caption{The energy spectrum of the end states with electron-electron interaction. There two electrons at the end in (a),(b) and (c), four electrons in (d) (e) and (f), six electrons in (g) (h) and (i). (a) (d) and (g) are corresponding to the Hubbard model with $U=1$. In (b) (e) and (h) a long range interaction are considered with $U=1,U_1=0.5$  and screened interaction with $U=1, U_1=0.5, \alpha=0.5$ are considered in (c) (f) and (i) }\label{pp}
\end{figure}
As illustrated in Fig.2 the first excited states are three-fold degenerate and there are level crossing at the $t=T$ in all cases. Combined with the result in the quantum well we conclude the degeneracy splitting is not an common case as claimed in Ref \onlinecite{Fu2006}. So this kind of 1D spin pumps are not ``$Z_2$ pump'' in general.\\
A $Z_2$ topological invariant is proposed to characterize the $Z_2$ pump by studying the time reversal polarization of the occupied bands. The theory of the time reversal polarization is a completely single-electron theory and no electron-electron interaction is considered in defining the $Z_2$ invariant. However, as claimed by the authors the $Z_2$ attribute of the spin pump is enforced by the electron-electron interaction. As discussed above if the single-electron picture remains valid the pump can not be a $Z_2$ pump. So it's the electron-electron interaction that make the spin pump ``$Z_2$''. So it's logically unsatisfactory to character a non-$Z_2$ pump by a $Z_2$ invariant within the single-electron theory and to use a single-electron invariant to character a $Z_2$ pump that inherently depends on the electron-electron interaction.\\
\begin{figure}
  \includegraphics[width=3.5in,clip=true]{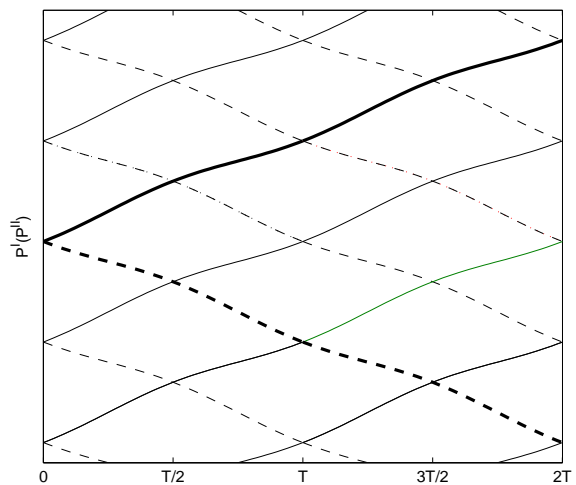}\\
  \caption{Continuous change of $P^I$ and $P^{II}$ in Ref \onlinecite{Fu2006}.}\label{pp}
\end{figure}
Now we discuss the time reversal polarization within the single-electron theory. The $Z_2$ invariant is defined by the continuous change of the time reversal polarization $P_\theta$ from $t=0$ to $t=T$. Though the topological invariant can be defined in this way to characterize the occupied bands, we must realize that the gauge invariant continuous change of $P_\theta$ as function of $t$ is not a $Z_2$ quantity as shown in Fig.3. The behavior of the system in the process of the adiabatic evolution is determined by the continuous change of the $P_\theta$. So the spin pump is not a $Z_2$ pump within the single-electron theory. We prefer to character the spin pump by a odd Chern number in this case\cite{Wu2013}.\\
As a final remark we discuss the relevance of this 1D spin pump to the spin Hall effect in 2D system. Only ``If the cylinder consists of a single unit cell in the circumferential (x) direction, then the magnetic flux threading the cylinder plays the role of the crystal momentum $k_x$ in band theory\cite{Fu07}.'' However, a cylinder consists of a single unit cell in the circumferential direction can not  be realized experimentally. So this pump can not be directly applied to the 2D topological insulator. The spin Hall effect must be explained by a real 2D pump\cite{Wu2013}.

\end{document}